\documentclass[pra,twocolumn,nofootinbib,floatfix,subeqn]{revtex4-1}
\usepackage[usenames]{color}
\usepackage{amsmath}
\usepackage{amssymb}
\usepackage{float}
\usepackage{graphicx}
\usepackage{xspace}
\usepackage{hyperref}

\newcommand{\bra}[1]{\ensuremath{\left\langle#1\right|}}
\newcommand{\ket}[1]{\ensuremath{\left|#1\right\rangle}}

\long\def\symbolfootnote[#1]#2{\begingroup%
\def\thefootnote{\fnsymbol{footnote}}\footnotetext[#1]{#2}\endgroup}

\begin{document}

\title{Single-atom imaging of fermions in a quantum-gas microscope}

%% Notice placement of commas and superscripts and use of &
%% in the author list
\author{Elmar Haller$^{1}$}
\author{James Hudson$^{1}$}
\author{Andrew Kelly$^{1}$}
\author{Dylan A. Cotta$^{1}$}
\author{Bruno Peaudecerf$^{1}$}
\author{Graham D. Bruce$^{1*}$}
\author{Stefan Kuhr$^{1\dag}$}

\date{22 June 2015}

\affiliation{
   $^1$University of Strathclyde, Department of Physics,\\
   Scottish Universities Physics Allicane (SUPA),   Glasgow G4 0NG, United Kingdom
}

\begin{abstract}
Single-atom-resolved detection in optical lattices using quantum-gas microscopes has enabled a new generation of experiments in the field of quantum simulation. While such devices have been realised with bosonic species, a fermionic quantum-gas microscope has remained elusive. Here we demonstrate single-site- and single-atom-resolved fluorescence imaging of fermionic potassium-40 atoms in a quantum-gas microscope setup, using electromagnetically-induced-transparency cooling. We detected on average 1000 fluorescence photons from a single atom within 1.5\,s, while keeping it close to the vibrational ground state of the optical lattice. A quantum simulator for fermions with single-particle access will be an excellent test bed to investigate phenomena and properties of strongly correlated fermionic quantum systems,
allowing for direct measurement of ordered quantum phases and out-of-equilibrium dynamics, with access to quantities ranging from spin-spin correlation functions to many-particle entanglement.
\end{abstract}

\maketitle
% INTRODUCTION

\symbolfootnote[1]{present address: University of St.\,Andrews, School of Physics and Astronomy, United Kingdom}
\symbolfootnote[2]{Electronic address: {\bf stefan.kuhr@strath.ac.uk}}

The ability to observe and control quantum systems at the single-particle level has revolutionised the field of quantum optics over the last decades. At the same time, the possibility to study quantum many-body systems in well-controlled engineered environments using ultracold atoms in optical lattices has proven to be a powerful tool for the investigation   of strongly-correlated quantum systems~\cite{Bloch:2008c}. It was only recently that the great challenge to have full single-site resolution and single-atom control in optical lattices was overcome by the development of quantum-gas microscopes~\cite{Bakr:2009,Sherson:2010}. These have been realized with bosonic $^{87}$Rb~\cite{Bakr:2009,Sherson:2010} and $^{174}$Yb atoms~\cite{Miranda:2014}. However, until this work and that presented in Refs.~\cite{Cheuk:2015,Parsons:2015}, single-site resolution for fermions has remained elusive.

Some of the most interesting effects in many-body quantum systems are due to the properties of strongly interacting Fermi gases. Within solid state physics, the fermionic nature of the electron is vital to understand a range of phenomena, such as electron pairing in superconductivity, and quantum magnetism (including colossal magneto-resistance). However, some of the properties of many-body fermionic quantum systems are very challenging to compute, even with the most advanced numerical methods, due to the antisymmetric nature of the wave function, and the resulting sign problem for quantum Monte-Carlo methods~\cite{Troyer:2005}. A quantum simulator for fermions with single-particle resolution would be an excellent test bed to investigate many of the phenomena and properties of strongly-correlated fermionic quantum systems. Such a fermionic quantum-gas microscope will provide the possibility to probe quantities that are difficult to access directly, such as spin-spin-correlation functions  or string-order~\cite{Endres:2011}. It would allow the study of out-of-equilibrium dynamics, the spreading of correlations~\cite{Cheneau:2012} and the build-up of entanglement in many-particle fermionic quantum systems~\cite{Pichler:2013}. It could perform quantum simulation of the Fermi-Hubbard model, which is conjectured to capture the key mechanism behind high-T$_{\rm c}$ superconductors,
 allowing researchers to gain insight into electronic properties that could potentially be applied in future materials engineering.

\begin{figure}[!h]
%\vspace{0.5cm}
    \begin{center}
        \includegraphics[width=\columnwidth]{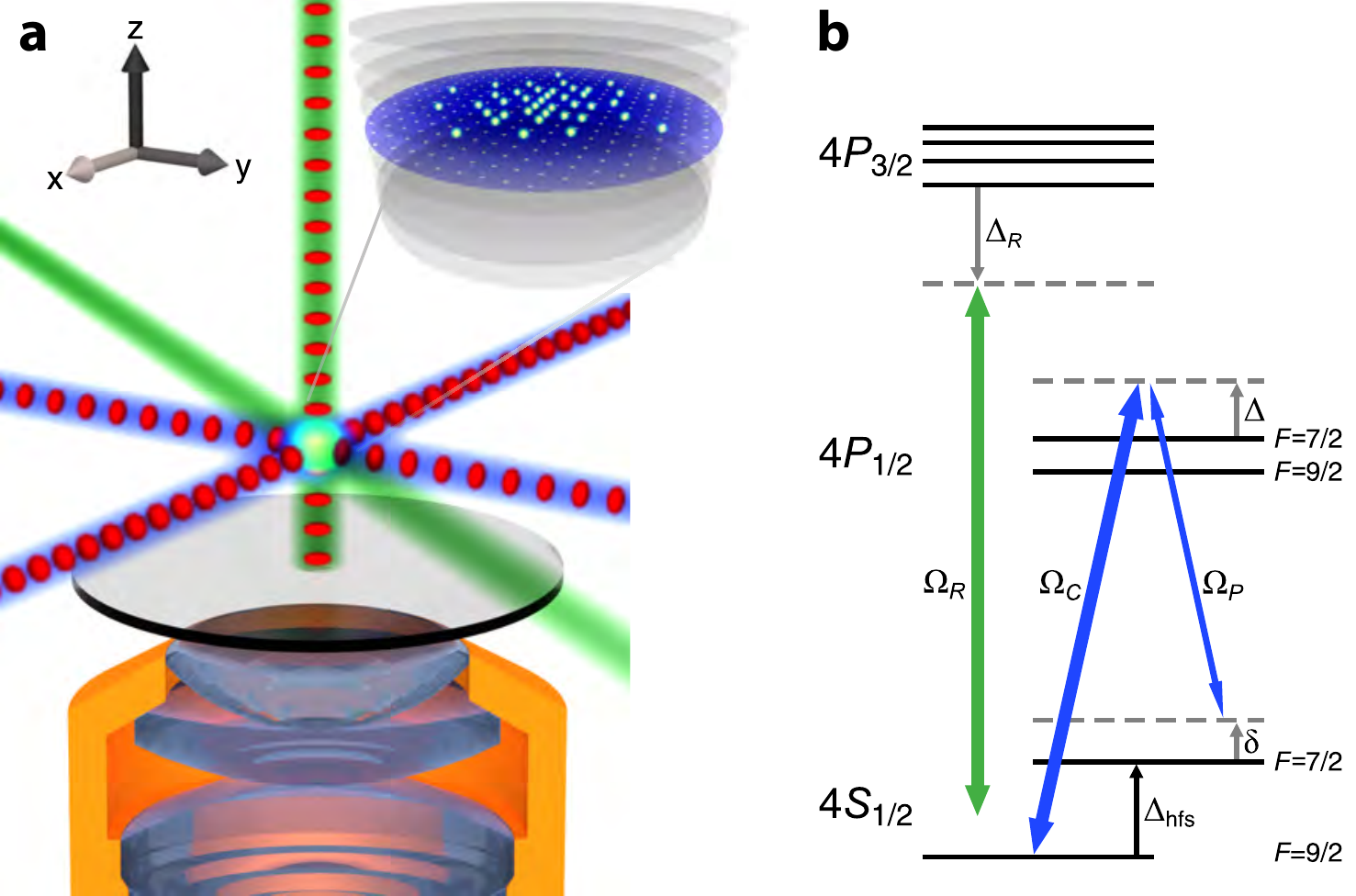}
    \end{center}
  \caption{{\bf Experimental setup, laser beam configuration and level scheme.}
 {\bf a,} Fermionic $^{40}$K atoms in an optical lattice were observed using fluorescence detection with a high-resolution optical microscope ($\mbox{NA} =0.68$, working distance $12.8$\,mm~\cite{Sherson:2010}). The optical lattice (red dots) is composed of two retro-reflected beams in the $x$- and $y$-axes, and a vertical beam retro-reflected from the coated vacuum window. Retro-reflected EIT cooling beams (blue) were overlapped with the horizontal optical lattice. Raman beams (green) were used to couple the motional states in the $z$-axis to the horizontal plane. Atoms were prepared in the focal plane of the microscope objective (inset). {\bf b,} Level scheme of the relevant states of $^{40}$K, with off-resonant Raman beams (green) and near-resonant (detuning $\Delta$) EIT coupling  and probe beams ($\Omega_C$ and $\Omega_P$, blue) with relative detuning $\Delta_{\rm hfs}+\delta$. \label{fig:setup}}
\end{figure}

\begin{figure*}
    \begin{center}
        \includegraphics[width=\textwidth]{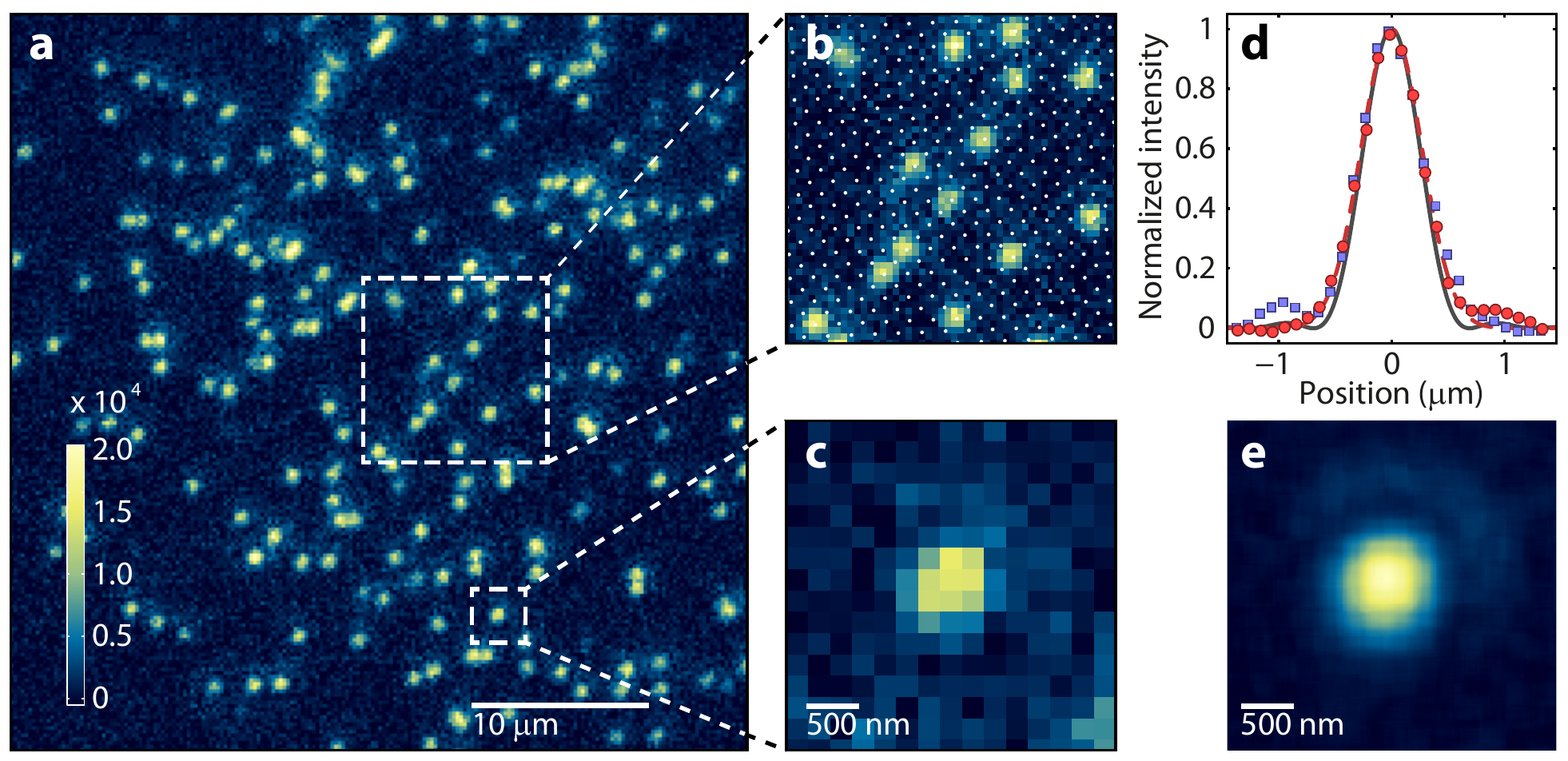}
    \end{center}
    \vspace{-0.5cm}
 \caption{{\bf Single-atom-resolved fluorescence images of fermions.}
    {\bf a,} Fluorescence image of a dilute cloud of $^{40}$K atoms in the optical lattice ($1.5$\,s exposure time with EIT cooling).
    {\bf b,} 10\,$\mu$m $\times$ 10\,$\mu$m subsection of {\bf a}; white dots mark the lattice sites.
    {\bf c,} Magnified subsection of {\bf a} showing an individual atom.
    {\bf d,} Horizontal (red circles) and vertical (blue squares) profiles through the centre of the averaged single-atom images {\bf e}, fitted (horizontal only) with a Gaussian profile (red dashed line). The grey line shows the diffraction-limited point-spread function (PSF) of the microscope objective.
    {\bf e,} Measured PSF using 640 single-atom images as presented in {\bf c}, averaged using subpixel shifting~\cite{Bakr:2009}. \label{fig:images}}
\end{figure*}

To achieve single-site-resolved detection of individual atoms on the lattice, it is desirable to maximise the fluorescence yield while at the same time maintaining a negligible particle loss rate and preventing the atoms from hopping between lattice sites. These conditions can be efficiently achieved by simultaneously laser cooling the atoms to sub-Doppler temperatures while detecting the fluorescence photons emitted during this process. However, cooling of fermionic alkaline atoms in optical lattices is challenging, as their low mass and small excited-state hyperfine splitting make it more difficult to obtain low temperatures using the standard technique of polarisation-gradient cooling.

In this work, we  achieved single-atom-resolved fluorescence imaging of $^{40}$K using electromagnetically-induced-transparency (EIT) cooling~\cite{Morigi:2000,Morigi:2003}. This technique has been used to efficiently cool trapped ions~\cite{Roos:2000} and to manipulate individual neutral atoms in an optical cavity~\cite{Muecke:2010,Kampschulte:2010,Kampschulte:2014}. EIT cooling is similar to grey molasses, recently used to cool atoms and enhance density in free-space atomic samples~\cite{Fernandes:2012,Grier:2013}.
Both methods rely on the existence of a spectrally narrow, Fano-like line profile~\cite{Lounis:1992} and dark resonances arising from quantum interference in a $\Lambda$-like level scheme.
In confining potentials with quantised vibrational levels, as is the case in our optical lattice, the narrow absorption line can selectively excite red-sideband transitions that cool the atomic motion by removing one vibrational quantum, while carrier and blue-sideband excitations are suppressed. We present EIT cooling as an alternative to Raman sideband cooling techniques~\cite{Hamann:1998}, which have been applied to image atomic clouds~\cite{Patil:2014} and individual atoms~\cite{Lester:2014}, and which have been used recently by other research groups to achieve single-site imaging of fermionic potassium~\cite{Cheuk:2015} or lithium~\cite{Parsons:2015}, independently of the work presented here.

%%%%%%%%%%%%%%%%%%%%%%%%%%%%%%%%%%%%%%%%%%%%%%%%%%%%%%%%%%%%%%%%%%%%%%%%%%
%BEGIN EXPERIMENT DESCRIPTION
%%%%%%%%%%%%%%%%%%%%%%%%%%%%%%%%%%%%%%%%%%%%%%%%%%%%%%%%%%%%%%%%%%%%%%%%%%

\begin{figure*}[!t]
    \begin{center}
        \includegraphics[width=\textwidth]{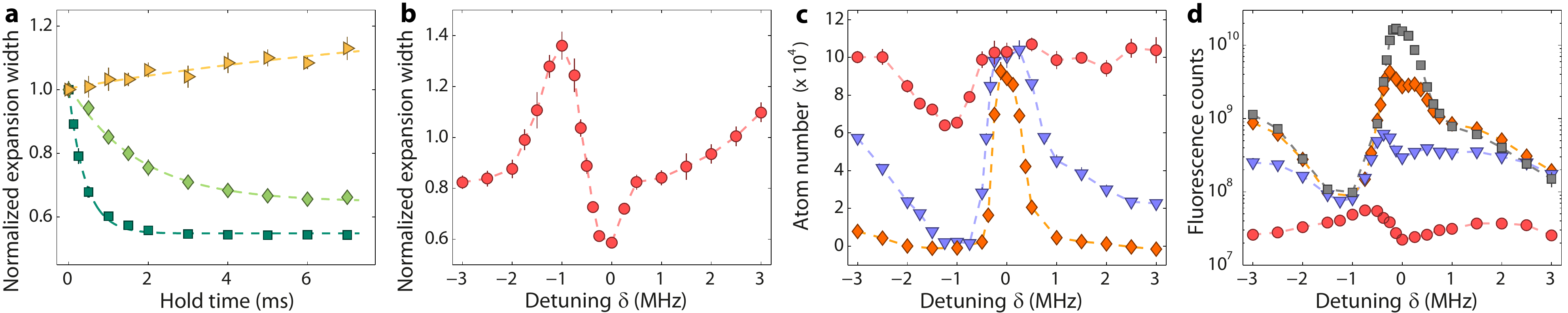}
    \end{center}
    \vspace{-0.5cm}
     \caption{{\bf Demonstration of EIT cooling.} {\bf a,} Expansion width of the atomic cloud after 1\,ms time-of-flight, normalised to the initial width, for varying duration of EIT cooling. Cooling was observed in the $y$-axis (green squares). In the $z$-axis, we observed cooling with Raman beams present (green diamonds) and heating without (yellow triangles). {\bf b,} Expansion width in $y$-axis as a function of two-photon detuning $\delta$, after 1.5\,ms of EIT cooling. {\bf c,} Atom numbers versus $\delta$, after $1.5$\,ms (red circles), $30$\,ms (blue triangles) and $300$\,ms (orange diamonds) of EIT cooling, measured with absorption imaging. {\bf d,} Integrated fluorescence counts over exposure times of $1.5$\,ms (red circles), $30$\,ms (blue triangles), $300$\,ms (orange diamonds) and $2000$\,ms (grey squares) during EIT cooling. Each datapoint in {\bf a} - {\bf d} is the average of five independent measurements, and the error bars represent the standard deviation.
    \label{fig:EIT}}
\end{figure*}

%%%%%%%%%%%%%%%%%%%%%%%%%%%%%%%%%%%%%%%%%%%%%%%%%%%%%%%%%%%%%%%%%%%%%%%%%%
%FIG 1
%%%%%%%%%%%%%%%%%%%%%%%%%%%%%%%%%%%%%%%%%%%%%%%%%%%%%%%%%%%%%%%%%%%%%%%%%%

The key component of our experimental setup is a high-resolution optical microscope~\cite{Sherson:2010} (Fig.\,\ref{fig:setup}a) with numerical aperture  $\mbox{NA}=0.68$, providing a diffraction-limited resolution of $580$\,nm (full width at half-maximum, FWHM) at the wavelength of the $4S_{1/2}\rightarrow 4P_{1/2}$ transition, $\lambda_{D1} = 770.1\,$nm ($D_1$-line).
We prepared a sample of cold fermionic $^{40}$K atoms within the focal plane of the microscope (Appendix) in a three-dimensional optical lattice (wavelength $\lambda=1064$\,nm). In order to implement the EIT-cooling scheme, we tuned both coupling and probe beams (Rabi frequencies $\Omega_C$ and $\Omega_P$) close to the $D_1$-line (see atomic level scheme in Fig.\,\ref{fig:setup}b). The coupling light was blue detuned by $\Delta = 10\,\Gamma$ ($\Gamma = 2 \pi \times 6.0$\,MHz is the linewidth of the $4P$ excited states) from the $F=9/2 \rightarrow F'=7/2$ transition in the lattice, and similarly, the probe beam was detuned by $\Delta = 10\,\Gamma$ from the $F=7/2 \rightarrow F'=7/2$ transition. Both pump and probe beams were derived from the same laser by passing the light through an electro-optical modulator (EOM), which generates sidebands at the frequency of the ground state hyperfine splitting, $\Delta_{\rm hfs}=2\pi \times 1.2858$\,GHz. A retroreflected EIT beam in $\sigma^+ - \sigma^-$ configuration is overlapped with each of the optical lattice beams in the $x$- and $y$-directions (Fig.\,\ref{fig:setup}a) to provide cooling in the horizontal plane. We avoided the use of vertical EIT beams to prevent stray light along the imaging axis. Instead, we used an off-resonant two-photon Raman transition red detuned by $\Delta_R=2\pi\times 8$\,GHz from the $4S_{1/2}\rightarrow 4P_{3/2}$ transition ($D_2$-line at $767.7$\,nm) to couple the motional states between the vertical and horizontal lattice axes (Appendix). One Raman beam was aligned vertically and the other one horizontally at a 45$^\circ$ angle between the $x$- and $y$-axes (Fig.~\ref{fig:setup}a). We blocked the stray light of the vertical Raman beam with narrowband interference filters (Alluxa) placed in front of the EMCCD camera, which detects fluorescence photons scattered from the EIT beams on the $D_1$ line.

%%%%%%%%%%%%%%%%%%%%%%%%%%%%%%%%%%%%%%%%%%%%%%%%%%%%%%%%%%%%%%%%%%%%%%%%%%
%  FIG 2
%%%%%%%%%%%%%%%%%%%%%%%%%%%%%%%%%%%%%%%%%%%%%%%%%%%%%%%%%%%%%%%%%%%%%%%%%%
Our cooling scheme allowed us to image individual $^{40}$K atoms in a sparsely populated thermal cloud (Fig.\,\ref{fig:images}a) with an  excellent signal-to-noise ratio. We detected on average $1000(400)$ fluorescence photons per atom within an illumination time of 1.5\,s using a lattice depth of $k_B\times 245(20)\,\mu$K on all axes (Appendix). Taking into account the collection efficiency of $8.6$\,\% of our imaging system (Appendix), we calculated a corresponding atomic fluorescence rate of $8(3)\times 10^3$~photons/s. This is large enough to unambiguously identify the presence or absence of an atom for each lattice site (Fig.\,\ref{fig:images}b and Appendix), however it is more than a factor of 10 smaller than those obtained in the quantum-gas-microscope setups with $^{87}$Rb~\cite{Bakr:2009,Sherson:2010}.
Our measured point spread function, obtained by averaging 640 fluorescence images of individual atoms (Fig.\,\ref{fig:images}e), has a FWHM of 630(10)\,nm, close to the diffraction-limited resolution of the microscope objective (Fig.\,\ref{fig:images}d).

%%%%%%%%%%%%%%%%%%%%%%%%%%%%%%%%%%%%%%%%%%%%%%%%%%%%%%%%%%%%%%%%%%%%%%%%%%
%  FIG 3
%%%%%%%%%%%%%%%%%%%%%%%%%%%%%%%%%%%%%%%%%%%%%%%%%%%%%%%%%%%%%%%%%%%%%%%%%%
In order to characterise the cooling process, we determined the cloud width for increasing duration of exposure to the EIT- and Raman beams. The cooling results in a decrease of the cloud width (measured using absorption images after rapid release from the lattice potential, see Appendix) to a stationary value, with a time constant of $0.42(2)$\,ms in the horizontal plane and $1.7(2)$\,ms in the vertical direction (Fig.\,\ref{fig:EIT}a). These time constants correspond to cooling rates of $6(1)\,\hbar \omega /$ms and $0.8(3)\,\hbar \omega /$ms respectively, where $\omega = 2\pi \times 300(12)$\,kHz is the spacing of vibrational levels in all three lattice axes. From our measured Rabi frequencies, $\Omega_C = 2\pi \times 4.8(1) $\,MHz, $\Omega_P = 2\pi \times 1.6(1)$\,MHz (see Appendix), we estimated the width of the narrow line in the Fano profile, $\gamma_+= \omega \Gamma / (4\sqrt{\Delta^2+\Omega_C^2+\Omega_P^2}) \approx  2\pi \times 6\,$kHz~\cite{Morigi:2003}. This value gives an upper limit for the combined scattering rate, $\gamma_+/2$, on the red sideband for both horizontal axes; our observed cooling (and fluorescence) rates are about a factor of three less. The cooling rate in the vertical direction is different as there are no vertical EIT beams, instead, cooling is achieved by transferring vibrational excitation quanta via Raman transitions from the  vertical axis to the cooled horizontal axes (Appendix). Without the Raman beams we observed no cooling in the vertical axis, but instead an increase of the vertical cloud width with time (Fig.\,\ref{fig:EIT}a), caused by heating due to spontaneously emitted photons from the EIT beams. We estimated a mean number of vibrational quanta of 0.9(1) in each the horizontal axis, after $5\,$ms of cooling, by measuring the steady state expansion energy with time-of-flight imaging (Appendix).

We studied the characteristics of EIT cooling in our experiment by measuring the cloud width, after $1.5$\,ms of cooling, as a function of the two-photon detuning $\delta$. On resonance, $\delta=0$, cooling works most efficiently (Fig.\,\ref{fig:EIT}b) as the atoms are pumped into the dark state and their lifetime is significantly enhanced (Fig.\,\ref{fig:EIT}c). However, at negative detuning, $\delta < 0$, we predominantly drive carrier and blue side-band transitions which heat the atoms (Fig.\,\ref{fig:EIT}b). For $\delta \neq 0$,  heating resulted in atom loss over timescales of milliseconds. Maximising the number of fluorescence photons for the single-atom imaging necessitates a compromise between a long lifetime (atoms in the dark state) and a large scattering rate. This can be seen in a shift of the maximum of fluorescence counts with increasing accumulation time (Fig.\,\ref{fig:EIT}d), from the position of strong heating ($\delta<0$) towards resonance ($\delta=0$), where most efficient cooling is achieved.
Note that Figs.\,\ref{fig:EIT}c and \ref{fig:EIT}d compare the atom number and the fluorescence counts of the complete cloud without layer preparation, in order to match the conditions of Figs.~\ref{fig:EIT}a and \ref{fig:EIT}b. However, the spatially varying light shift due to the Gaussian intensity profile of the lattice beams causes a position-dependent detuning $\Delta$. As a consequence, optimum conditions for EIT cooling are only met in the central $70\times70$ lattices  sites, where $\Delta$ changes by less than $20\,$MHz. This fact causes loss of atoms outside of the area of most efficient cooling, resulting in a reduced lifetime for the full cloud of 6.6(5)\,s. The lifetime of the atoms inferred by measuring fluorescence from the central $70\,\times\,70$ lattice sites is 30(4)\,s (see Appendix, Fig.~\ref{fig:LIFE}).

%%%%%%%%%%%%%%%%%%%%%%%%%%%%%%%%%%%%%%%%%%%%%%%%%%%%%%%%%%%%%%%%%%%%%%%%%%
%FIG4
%%%%%%%%%%%%%%%%%%%%%%%%%%%%%%%%%%%%%%%%%%%%%%%%%%%%%%%%%%%%%%%%%%%%%%%%%%

\begin{figure}[!t]
   \begin{center}
        \includegraphics[width=0.9\columnwidth]{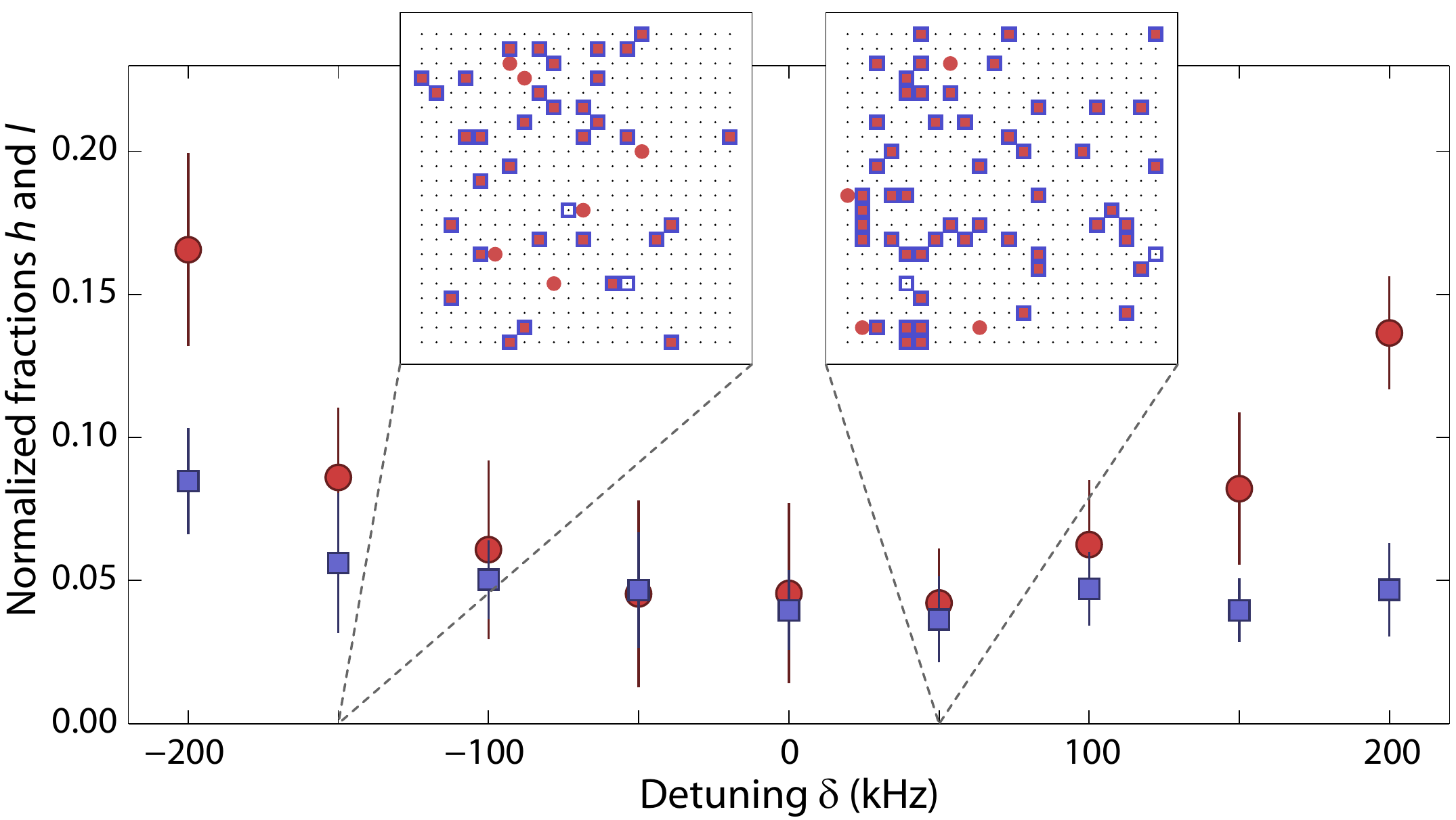}
            \end{center}
        \vspace{-0.5cm}
   \caption{{\bf Hopping and atom losses.} Fraction of atoms, $h$, appearing (blue squares) on a previously empty site in the second of two successive fluorescence images (500\,ms exposure time, 500\,ms delay between the two images), normalized to the number of atoms in the first image. Also shown is the fraction of atoms, $l$, that has disappeared between the two images (red dots). The panels show binary maps of the lattice site occupations of the first (red circles) and second image (blue squares).
    They show regions of $20\times20$ lattice sites, however the data is analysed in a region of $70\times70$ sites. Each data point results from the average of ten fluorescence images; the error bars represent the standard deviation.\label{fig:hopping}}
\end{figure}

We investigated in more detail to what extent the atoms are lost and move in the lattice during the imaging process. For this purpose, we recorded two successive fluorescence images of the same atoms with 500\,ms exposure time each, and a 500\,ms separation between the two images. From each of the images we determined the atom distribution on the lattice using a deconvolution algorithm (Appendix), providing us with a binary map of the lattice site occupation. To estimate hopping, we calculated the fraction of atoms, $h$, that appeared on previously empty lattice sites as well as the fraction of atoms lost, $l$,  from the first to the second image (Fig.\,\ref{fig:hopping}). Varying $\delta$, we measured hopping fractions as low as $h=3.5(1.5)\,\%$ and observed that $l$ increased from $4.2(1.9)\,\%$ at $\delta=0$ to larger values for $\delta \neq 0$. We attribute this increased loss to the less efficient cooling due to the absence of a dark state, as discussed above (Fig.\,\ref{fig:EIT}c). Even at $\delta=0$, both loss and hopping could result from off-resonant excitation by the EIT- and Raman beams to the $4P_{1/2}$ and $4P_{3/2}$ excited states.
Both $4P$ states experience a light shift more than five times greater and opposite to the ground-state light shift, due to the proximity of  the lattice wavelength to the $4P\rightarrow3D$ resonance at 1180\,nm. In our case, the use of blue detuned EIT beams ensures that the atoms preferentially scatter off a trapped dressed state. However, an atom resonantly excited to the $4P$ state experiences a repulsive potential which causes the excited-state wavepacket to expand rapidly in space. During the subsequent spontaneous emission the atoms can decay into a high vibrational state or a high band in which the atoms are no longer bound to the lattice sites. These hot atoms can now move across the lattice and then potentially be recooled and bound to a different lattice site. We are currently investigating this heating process in more detail using numerical simulations of the full quantum evolution of an atom on a few lattice sites.

%%%%%%%%%%%%%%%%%%%%%%%%%%%%%%%%%%%%%%%%%%%%%%%%%%%%%%%%%%%%%%%%%%%%%%%%%%
%OUTLOOK
%%%%%%%%%%%%%%%%%%%%%%%%%%%%%%%%%%%%%%%%%%%%%%%%%%%%%%%%%%%%%%%%%%%%%%%%%%
%

The single-site-resolved detection of $^{40}$K in an optical lattice will open the path to the study of strongly correlated fermionic quantum systems with unprecedented insight into their properties.
In particular, our new technique will allow the study of strongly correlated phases in the Fermi-Hubbard model, e.g., counting thermal excitations in fermionic Mott insulators~\cite{Joerdens:2008a,Schneider:2008a,Joerdens:2010} and directly probing local entropy distributions~\cite{Sherson:2010}, or detecting the propagation of correlations after quenches~\cite{Cheneau:2012}. The quantum-gas microscope will also enable manipulation of the system on a local scale with single-site addressing resolution~\cite{Weitenberg:2011}, allowing  deterministic preparation of non-thermal initial states~\cite{Fukuhara:2013} and probing the ensuing dynamics. This would allow, for example, further investigation of intriguing phenomena in Luttinger liquid theory, such as spin-charge separation~\cite{Kollath:2005} and  growth of many-body spatial entanglement~\cite{Pichler:2013}.
The high-resolution microscope could also be used to implement proposals for removing high-entropy regions by locally modifying the confining potential \cite{Bernier:2009} in order to attain low-entropy states. The fermionic quantum-gas microscope is an ideal tool to detect the onset of antiferromagnetic ordering~\cite{Snoek:2008,Paiva:2010,Simon:2011,Greif:2013,Hart:2015} with single-site resolution, by directly measuring the spin-spin correlations.  In different lattice geometries these measurements would allow the study of, e.g., both equilibrium properties and out-of-equilibrium dynamics in frustrated spin systems~\cite{Eisert:2015}.

\section*{Acknowledgements}
 We thank G. Morigi, A. Daley, and A. Buyskikh for fruitful discussions. We acknowledge the contribution of A. Schindewolf, N. Sangouard and J. Hinney during the construction of the experiment. We acknowledge support by EU (ERC-StG FERMILATT, SIQS, Marie Curie Fellowship to E.H.), EPSRC, Scottish Universities Physics Alliance (SUPA).

%% Put the bibliography here, most people will use BiBTeX in
%% which case the environment below should be replaced with
%% the \bibliography{} command.
%\bibliographystyle{natbib}
\bibliography{SingleFermionsBib}

%%%%%%%%%%%%%%%%%%%%%%%%%%%%%%%%%%%%%%%%%%%%%%%%%%%%%%%%%%%%%%%%%%%%%%%%%%
%METHODS
%%%%%%%%%%%%%%%%%%%%%%%%%%%%%%%%%%%%%%%%%%%%%%%%%%%%%%%%%%%%%%%%%%%%%%%%%%

\section*{Appendix}
\subsection*{Loading of $^{40}$K atoms into the optical lattice}
Our experimental sequence started with an atomic beam of $^{40}$K  produced by a two-dimensional magneto-optical trap (MOT) loaded from $10^{-7}$ mbar of potassium vapour (our atomic source is enriched to $3\,\%$ $^{40}$K). This atomic beam passes through a differential pumping section to a standard three-dimensional MOT, which captures $2 \times 10^{8}$ atoms in 4s.
The atoms were further cooled in an optical molasses and loaded into an optical dipole trap formed by two beams of 100\,W each (beam radius $w_0 = 150\,\mu$m, wavelength 1070\,nm) crossing at an angle of 15$^{\circ}$. The atoms were then loaded into a translatable optical tweezer (beam radius $w_0=50\,\mu$m, 18\,W power) and moved 13\,cm within 1.5\,s to within the field-of-view of the high-resolution microscope~\cite{Sherson:2010}. We initially transferred $1\times 10^6$  $^{40}$K  atoms from the single-beam trap into a crossed optical dipole trap which is formed by the optical lattice beams with their retro-reflectors blocked by mechanical shutters and evaporatively cooled the atoms to $T=4.0(1)\,\mu$K. We then transferred the sample into the vertical optical lattice (beam radius $w_0 = 60\,\mu$m) and subsequently switched on both horizontal optical lattices ($w_0 = 75\,\mu$m) with the retro-reflectors unblocked. The lattice depth was determined by measuring the frequency of the blue sideband transitions in the lattice using Raman transitions. Driving this sideband results in an increase of the cloud width after time-of-fight. The error of the lattice depth cited in the main text is the $1/e$-width of the excitation peak of the spectrum.

\subsection*{Preparation of 2D atomic samples}
In order to prepare a thin sample within the focal plane of the microscope, we used a position-dependent microwave transition in a vertical magnetic field gradient~\cite{Sherson:2010} of $\partial B/\partial z = 80$\,G/cm. This gradient results in a shift of the ground state hyperfine transition $\left|F=9/2, m_F=-9/2\right> \rightarrow \ket{F=7/2, m_F=-7/2}$ of $20$\,kHz/$\mu$m. We initially optically pumped the atoms into the $\ket{F=9/2, m_F=-9/2}$ state while holding them in a shallow 3D optical lattice, before applying a microwave sweep (HS1-pulse~\cite{Weitenberg:2011} of 5\,kHz width) to transfer atoms out of one selected layer into the $\ket{F=7/2, m_F=-7/2}$ state. All atoms remaining in $F=9/2$ were then removed using a circularly polarised resonant laser beam on the $F=9/2 \rightarrow F'=11/2$ cycling transition of the $D_2$ line. The microwave transfer and removal processes have an overall efficiency of $>$90\,\%, resulting in residual atoms in the $F=7/2$ manifold across all layers. In order to remove these atoms, we optically pumped the atoms back to the initial $\ket{F=9/2, m_F=-9/2}$ state and again performed a microwave transfer followed by a removal pulse. Using this technique, we were able to prepare about 100 atoms in a single layer of the three-dimensional optical lattice, with a negligible population in other layers. Note that for the images in Fig.\,\ref{fig:images}, we prepared two adjacent layers within the depth of focus of the imaging system, in order to increase the number of atoms visible.
For the datasets presented in Fig.~\ref{fig:EIT}, we used absorption images of the full atomic cloud, without single-layer preparation. The atoms are initially in $F=9/2$, but once the EIT cooling beams are switched on, the population is predominantly in state $F=7/2$.

\subsection*{Analysis of fluorescence images}
In order to analyse our fluorescence images, we used  procedures and computer algorithms based on earlier work~\cite{Sherson:2010}. The precise orientation of the lattice angles, $47.599(3)^\circ$ and $-42.833(3)^\circ$ with respect to the horizontal baseline of the Figs.\,\ref{fig:images}a and \ref{fig:images}b, and the magnification (5.133 pixels on the CCD camera correspond to $\lambda/2$ at the atoms; $\lambda=1064\,$m is the lattice wavelength) were determined by evaluating the mutual distances between 600 isolated atoms from several fluorescence images (see Supplementary Information of Ref.~\cite{Sherson:2010}). Lattice angles and spacings were taken as the input for our deconvolution algorithm~\cite{Sherson:2010}, which uses the averaged point spread function from Fig.\,\ref{fig:images}e to determine the occupation of the lattice sites. To estimate the average fluorescence rate, we summed the number of counts from a $3\times3$ pixels area centered on each occupied lattice site, and converted this value to a photon number. The resulting histogram of the detected photon number follows a Gaussian distribution of mean 1000 and standard deviation of 400. The same analysis performed on the empty lattice sites yielded a photon number of less than 200.

Our imaging system has a total fluorescence collection efficiency of $8.6$\,\%, given by the objective's solid angle of $\Omega / (4\pi) = 13$\,\%, the 80\,\% transmission through the objective, lenses and interference filters and the 83\,\% quantum efficiency of the EMCCD camera (Andor iXon 897).

\subsection*{Characterisation of EIT cooling\\ using time-of-flight imaging}
We determined the kinetic energy of a cloud of atoms in time-of-flight measurements after a rapid release from the lattice potential. The initial change in kinetic energy due to the EIT- and Raman beams was different for the horizontal and vertical axes, with a horizontal cooling rate of $83(14)\,\mu$K/ms during the first 150$\,\mu$s and a vertical cooling rate of $12(4)\,\mu$K/ms during the first $400\,\mu$s. After a hold time of $5\,$ms with EIT cooling, the atoms maintained constant horizontal and vertical expansion energies of $10.0(2)\,\mu$K and $14.8(2)\,\mu$K.  These kinetic energies correspond to mean excitation numbers of $0.9(1)$ and $1.6(1)$, respectively, for a harmonic oscillator potential with a trap frequency $\omega=2\pi \times 300(12)\,$kHz.\\

We determined the combined Rabi frequencies of all EIT beams by measuring the off-resonant scattering rate of one beam, with the EOM switched off. The coupling and probe beam Rabi frequencies, $\Omega_C = 2\pi \times 4.8(1)$\,MHz and $\Omega_P = 2\pi \times 1.6(1)$\,MHz, are determined by the modulation depth of the EOM (25\,\% of the power in the first sideband) and by taking into account the transition strengths. The errors of the Rabi frequencies do not take into account the fact that there could be different distributions of the magnetic sublevels during this measurement compared to EIT cooling in the lattice.\\

\subsection*{Raman coupling of motional axes}
In our setup, we achieved cooling of the vertical axis by transferring its vibrational excitation quanta, via Raman transitions, to the cooled horizontal axes~\cite{Steinbach:1997}. We used two Raman beams of identical frequency, such that the exchange of vibrational quanta between two axes is a resonant two-photon process. The Rabi frequency for the transition $\ket{n_x,n_z+1}\rightarrow\ket{n_x+1,n_z}$ can be calculated as $\Omega_R=\Omega^{0}\bra{n_x+1,n_z}e^{i\delta\mathbf{k}\cdot\mathbf{r}}\ket{n_x,n_z+1}$, where $n_x$ and $n_z$ are the vibrational quantum numbers,   $\delta\mathbf{k}=\mathbf{k_1}-\mathbf{k_2}$ is the momentum transfer from the Raman beams, and  $\Omega^{0}$ is the two-photon Rabi frequency of the carrier transition.  Expanding the exponential yields $\Omega_R=\Omega^{0}\eta_x\eta_z\sqrt{(n_x+1)(n_z+1)}$, where we have used the Lamb-Dicke factors $\eta_j=\mathbf{\delta k}\cdot\mathbf{e_j} \Delta x_0^j$ (for our parameters, $\eta_x = 0.118(2)$, $\eta_z=0.168(3)$, with $\Delta x_0^j=\sqrt{\hbar/(2m\omega)}=20.5(4)$\,nm the width of the ground state wavepacket of the harmonic oscillator in directions $j=x,z$ ($m$ is the mass of a $^{40}$K atom). We estimate a Rabi frequency of $\Omega_R = 2\pi\times27(2)$\,kHz by measuring how the vertical expansion width after 0.5\,ms of cooling changes as a function of the detuning between the two Raman beams.\\

\subsection*{Lifetime measurements}
\begin{figure}[[!h]
      \begin{center}
        \includegraphics[width=0.90\columnwidth]{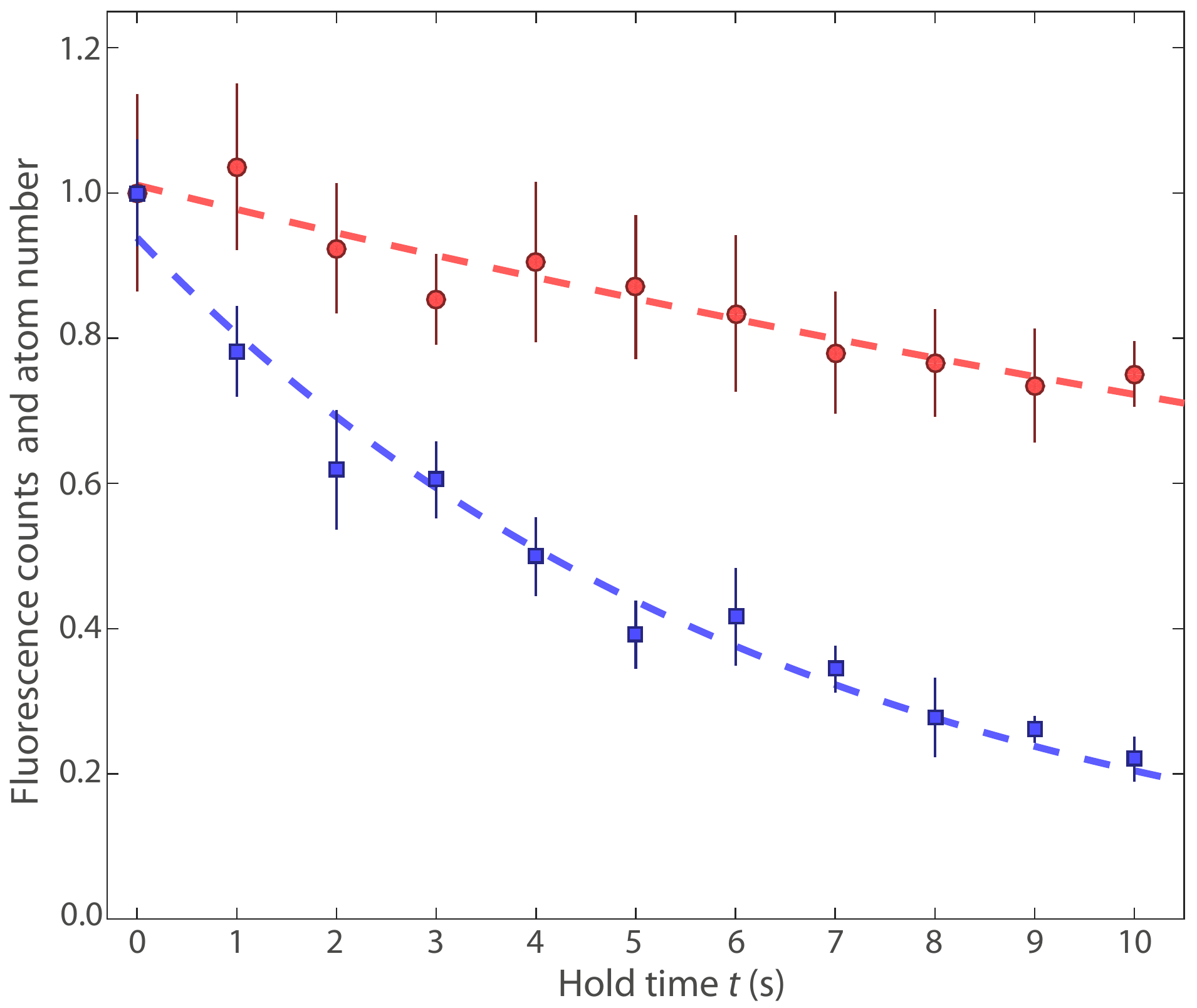}
                   \end{center}
    \caption{{\bf Lifetime measurement.} The curves show fluorescence counts within a region of $70\times70$ lattice sites integrated over a period of 250\,ms (red circles) and atom number of the complete cloud measured by absorption imaging (blue squares) for varying duration, $t$, of EIT cooling. Both curves were normalised to the respective initial values at $t=0$. We determine the time constants of 30(4)\,s and 6.6(5)\,s from exponential fits (dashed lines). \label{fig:LIFE}}
\end{figure}

\end{document}